\documentclass[aps,pra,twocolumn,amsmath,amssymb,superscriptaddress,longbibliography]{revtex4-1}
\UseRawInputEncoding
\usepackage[english]{babel}
\usepackage[english]{babel}
\usepackage{amssymb}
\usepackage{dcolumn}
\usepackage{bm}
\usepackage{graphicx}
\usepackage{amsmath}
\usepackage{graphicx}       
\usepackage{mathrsfs}                             
\graphicspath{{pict/}{}}
\usepackage{dcolumn}
\usepackage{bm}
\usepackage[dvipdfm, pdfstartview=FitH, CJKbookmarks=true, bookmarksnumbered=true, bookmarksopen=true, colorlinks=true, pdfborder=001, citecolor=blue, linkcolor=blue, linktocpage=true] {hyperref}
\begin{document}

\title{Tunable symmetry-protected higher-order topological states with fermionic atoms in bilayer optical lattices}

\author{Zhoutao Lei}
\affiliation{Guangdong Provincial Key Laboratory of Quantum Metrology and Sensing $\&$ School of Physics and Astronomy, Sun Yat-Sen University (Zhuhai Campus), Zhuhai 519082, China}
\affiliation{State Key Laboratory of Optoelectronic Materials and Technologies, Sun Yat-Sen University (Guangzhou Campus), Guangzhou 510275, China}

\author{Linhu Li}
\email{lilh56@mail.sysu.edu.cn}
\affiliation{Guangdong Provincial Key Laboratory of Quantum Metrology and Sensing $\&$ School of Physics and Astronomy, Sun Yat-Sen University (Zhuhai Campus), Zhuhai 519082, China}

\author{Yuangang Deng}
\email{dengyg3@mail.sysu.edu.cn}
\affiliation{Guangdong Provincial Key Laboratory of Quantum Metrology and Sensing $\&$ School of Physics and Astronomy, Sun Yat-Sen University (Zhuhai Campus), Zhuhai 519082, China}

\date{\today}

\begin{abstract}
Higher-order topological states that possess gapped bulk energy bands and  exotic topologically protected boundary states with at least two dimension lower than the bulk have significantly opened a new perspective for understanding of topological quantum matters. Here, we propose to generate two-dimensional topological boundary states for implementing synthetic magnetic flux of ultracold atoms trapped in bilayer optical lattices, which includes Chern insulator, Dirac semimetals, and second-order topological phase (SOTP) by the interplay of the two-photon detuning and effective Zeeman shift. These observed topological phases can be well characterized by the energy gap of bulk, Wilson loop spectra, and the spin textures at the higher symmetric points of system. We show that the SOTP exhibits a pair of $0$D boundary states. While the phases of Dirac semimetals and Chern insulator support the conventional $1$D boundary states due to the principle of bulk-boundary correspondence. Strikingly,  the emerged boundary states for Dirac semimetals and SOTP are topologically protected by $\cal P T$-symmetry and chiral-mirror symmetry ($\mathcal{\widetilde{M}}_{\alpha}$), respectively. In particular, the location of $0$D corner states for SOTP which are associated with existing $\mathcal{\widetilde{M}}_{\alpha}$-symmetry can be highly manipulated by tuning magnetic flux. Our scheme herein provides a platform for emerging exotic topological boundary states, which may facilitate the study of higher-order topological phases in ultracold atomic gases.
\end{abstract}

\maketitle

\section{introduction\label{Sec1}}
Topological quantum matters, which are characterized by the nontrivial topological invariant and exhibit fundamentally physical phenomena with versatile applications, have led to tremendous advances in recent years~\cite{RevModPhys.82.1959,RevModPhys.82.3045,RevModPhys.83.1057,RevModPhys.88.035005}. An important physical consequence of their topological nontrivial bulk bands is hosting exotic topologically protected boundary states. In particular, higher-order topological phases which is beyond the conventional bulk-boundary correspondence principle have attracted much attention in recent years~\cite{Benalcazar61,PhysRevB.96.245115,PhysRevLett.119.246401,PhysRevLett.119.246402,PhysRevLett.120.026801,PhysRevB.97.205135,PhysRevB.97.241405,PhysRevB.98.205147,PhysRevLett.122.076801,PhysRevLett.123.016805,PhysRevLett.123.016806,PhysRevX.9.011012,PhysRevLett.123.167001,PhysRevLett.123.186401,PhysRevLett.123.256402,PhysRevLett.125.166801,LI2021,2206.11296}.
In these significant advances, a $d$-dimensional ($d$D) $n$th-order topological phases will support the topological boundary states localized in their $(d-n)$D boundaries. The conventional topological states including topological insulators and topological semimetals, which possess the edge states with one dimension lower than the bulk, are classified as first-order topological states with $n=1$. To date, exotic higher-order topological phases have been realized in a wide range of physical systems, including solid materials~\cite{Schindler2018,Kempkes2019,acs.nanolett.9b00844}, photonic and phononic crystals~\cite{Noh2018,Li2018,PhysRevLett.122.233902,PhysRevLett.122.233903,ElHassan2019,Serra-Garcia2018,Xue2019,Ni2019,Zhang2019,PhysRevLett.124.206601,Zhang2020}, as well as in microwave and electric circuits~\cite{Peterson2018,Imhof2018}. To take advantage of the rich boundary physics of topological matters, it is necessary to explore the interplay of boundary states with distinct properties including their dimensions and locations~\cite{PhysRevB.100.075415,PhysRevLett.124.227001,PhysRevLett.125.017001,PhysRevLett.125.126403,PhysRevB.102.041122,PhysRevLett.126.156801,PhysRevLett.127.066801}.

Meanwhile, the currently available techniques of realizing spin-orbit (SO) coupling~\cite{NATYJ2011SOCAT,PRLZW2012SOCAT,Ji2014,SCPJW2016SOC,NSHL2016} and Raman-assisted tunneling~\cite{PhysRevLett.107.255301,PhysRevLett.111.185301,PhysRevLett.111.185302,NaJG2014,NPAM2014,NPKCJ2015,science.1259052} for ultracold quantum gases have provided a new platform for exploring exotic topological quantum matters in a clean environment and controllable way~\cite{RevModPhys.83.1523,Goldman_2014,Zhai_2015,Topologicalcoldatoms,RevModPhys.91.015005}. Of particular interest, 1D Su-Schrieffer-Heeger model~\cite{Atala2013}, 2D Haldane model~\cite{Jotzu2014}, and 3D topological semimetals~\cite{Song20191,Wang271} have been experimental realized for cold atoms in optical lattices. In addition, these novel topological phases for ultracold quantum gases can be characterized by measuring the closing and opening of the bulk gap via the
Landau-Zener transition~\cite{Jotzu2014}, Chern number of bands~\cite{NPAM2014,science.1259052}, Bloch state tomography ~\cite{science.aad5812,science.aad4568}, and quantum quenches~\cite{PhysRevB.88.104511,PhysRevLett.113.076403,PhysRevLett.115.236403,D¡¯Alessio2015,PhysRevLett.121.090401,ZHANG20181385,Songeaao4748}. In their pioneer explorations, the realization of these topological quantum phases is focusing on conventional first-order topological phases. The higher-order topological states remains a challenging task to synthesize, despite some recent theoretical advances in interacting systems~\cite{PhysRevLett.123.060402,PhysRevA.100.023602,PhysRevB.99.020508,PhysRevA.103.013307}, non-Hermitian system~\cite{PhysRevLett.123.073601}, and orbital angular momentum mediated-cold atoms~\cite{PhysRevB.100.205109}. Based on currently available experimental techniques for cold atoms~\cite{NATYJ2011SOCAT,PRLZW2012SOCAT,Ji2014,SCPJW2016SOC,NSHL2016, PhysRevLett.107.255301,PhysRevLett.111.185301,PhysRevLett.111.185302,NaJG2014,NPKCJ2015,NPAM2014,science.1259052}, an interesting question is that whether higher-order topological states with tunable topologically protected boundary states at different dimensions and locations can be realized with the help of advantages of cold atoms. An achievable experimental proposal as well as a simple accessible measurement method will significantly motivate the relevant studies of novel higher-order topological states in ultracold quantum gases~\cite{PhysRevLett.123.060402,PhysRevLett.123.073601}.

In this work, we propose an experimental scheme to realize tunable symmetry-protected second-order topological phase (SOTP) for Raman-assisted SO coupled ultracold quantum gases trapped in a bilayer optical lattice. Due to the interplay of tunable two-photon detuning and Zeeman shift induced by a gradient magnetic field, the system possesses chiral-mirror symmetry ($\mathcal{\widetilde{M}}_{\alpha}$) protected SOTP and $\cal P T$-symmetry protected Dirac semimetals. We show that the gapped topological phases can be well distinguished by calculating the  Wilson loop spectra and spin texture at the higher symmetric points of system. Of particular interest, the $0$D corner states of SOTP with respect to $\mathcal{\widetilde{M}}_{\alpha}$ symmetry are observed, which is essentially different from the first-order topological states hosting the principle of bulk-boundary correspondence. Furthermore, our system holds distinctly $\mathcal{\widetilde{M}}_{\alpha}$ symmetries by employing the different quantized synthetic magnetic flux. The proposed scheme has the advantage that both the dimensions and locations of symmetry-protected topological boundary states can be highly controlled by tuning the magnetic flux in experiments, which may provide new insights in understanding exotic higher-order topological states and facilitate their experimental detection for ultracold quantum gases.

This paper is organized as follows. In Sec~\ref{Sec2}, we introduce our model and Hamiltonian for fermionic atoms trapped in bilayer lattices. Section~\ref{Sec3} is devoted to study the topological quantum phases and map the phases diagram of system. In Sec~\ref{Sec4}, we discuss the  first-order topologically protected $1$D boundary states for Dirac semimetals and CI. In Sec~\ref{Sec5}, we present the exotic bulk-edge correspondence for SOTP. Finally, a brief summary is given in Sec~\ref{Sec6}.

\section{Model and Hamiltonain\label{Sec2}}

\begin{figure}[!htp]
\includegraphics[width=1\columnwidth]{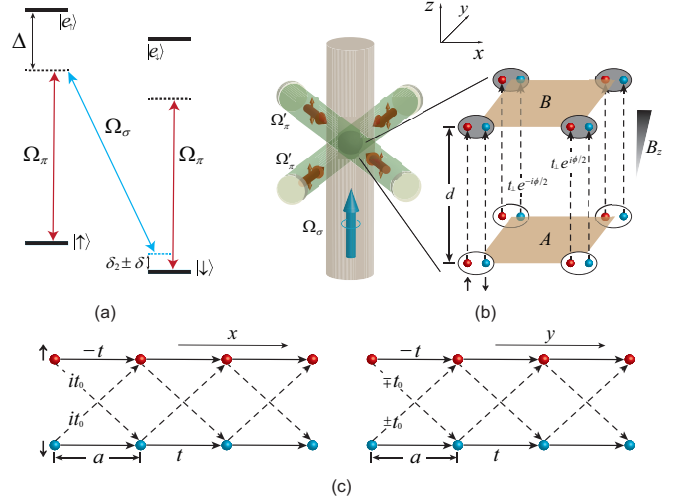}
\caption{\label{model}(color online). (a) Level diagram of a $\Lambda$ atomic system. (b) Proposed Bilayer optical lattice and laser configuration for generating Raman-assisted hoping. (c) The illustration of intralayer staggered Raman-assisted spin-flip hopping along the physical dimensions $x$- and $y$-directions.}
\end{figure}

We consider an ultracold gas of $N$ fermionic atoms subjected to a gradient magnetic field ${\bf B}=b_0z\hat{z}$ along the quantization $z$-axis. Figure~\ref{model}(a) displays the atomic level structure, which includes two electronic ground states $|\uparrow\rangle$ and $|\downarrow\rangle$, and one excited state $|e_{\uparrow}\rangle$. The magnetic quantum numbers of these electronic states satisfy $m_{\uparrow}=m_{e_{\uparrow}}$ and $m_{\uparrow}=m_{\downarrow}-1$. Then the fermionic atoms are deep confined by a red-detuned $2$D bilayer square optical lattices $\mathcal{U}(\mathbf{r})=-U_{ol}[\cos^2(k_Lx)+\cos^2(k_Ly)]$. Here $U_{ol}$ is the depth of optical lattice and $k_L$ is the wave vector of lattice laser with $a=\pi/k_L$ being the lattice constant. The two layers optical lattices are spatial separated along $z$-direction with distance $d$, as displayed in Fig.~\ref{model}(b). Recently, the creation of bilayer optical lattice has been predicted in various theoretical proposals~\cite{PhysRevA.100.053604,PhysRevLett.125.030504,PhysRevLett.126.103201} and experimental realization~\cite{meng2021atomic}. Specifically, the atomic transition from the ground state $|\uparrow\rangle$ to the excited state $|e_{\uparrow}\rangle$ is driven by a pair of $\pi$-polarized standing-wave lasers in $x$-$y$ plane, corresponding to Rabi frequencies $\Omega'_{\pi}\sin(k_Lx-k_Ly)$ and $i\Omega'_{\pi}\sin(k_Lx+k_Ly)$, respectively. Then the spatially dependent total Rabi frequency is $\Omega_{\pi}[\sin(k_Lx)\cos(k_Ly)+i\cos(k_Lx)\sin(k_Ly)]$ with $\Omega_{\pi}=\sqrt{2}e^{i\pi/4}\Omega'_{\pi}$.

To generate SO coupling, the atomic transition $|\uparrow\rangle\leftrightarrow |e_{\downarrow}\rangle$ is driven by a $\sigma$-polarized plane-wave laser propagating along $z$-direction with Rabi frequency $\Omega_{\sigma}$ ($\Omega_{\sigma}e^{i\phi}$) for atoms in layer A (B). We should note that the tunable phase difference $\phi=k_Ld$ is determined by the distance of the bilayer optical lattice. In addition, the tunable Zeeman shift difference between layers A and B is $\delta=g_F \mu_Bb_0 d$ induced by the gradient magnetic field, where $g_F$ is Land\'{e} $g$-factor of the hyperfine state and $\mu_B$ is the Bohr magneton. As a result, the Zeeman splittings between $|\uparrow\rangle$ and $|\downarrow\rangle$ states for layers A and B are $\delta_2+\delta$ and  $\delta_2-\delta$, as shown in Fig~\ref{model}(a). Indeed, the large Zeeman shift $\delta$ can also inhibit the spin-independent tunneling between two layers of cold atoms~\cite{PhysRevLett.111.185301,PhysRevLett.111.185302}. To create the Raman-assisted spin-flipping, the atoms are also illuminated by another Raman laser coupling bilayer lattices of cold atoms, corresponding to the spatial-independent Rabi frequency $\Omega_{\bot}$. As we shall see below, this spin-flipping hopping will play an important role in realizing second-order topological boundary states.

In the large atom-light detuning $\Delta$, the excited state $|e_{\uparrow}\rangle$ can be adiabatically eliminated. Then the single-particle Hamiltonian of the atom-light system reads
\begin{eqnarray}\label{Eq.realspace}
{\boldsymbol{h}}=&&\frac{{\mathbf{p}}^{2}}{2M}+M_x\mathbf{(r)}\hat{\sigma}_{x}(\hat{\tau}_0+
\hat{\tau}_z)/2+M_y\mathbf{(r)}\hat{\sigma}_{y}(\hat{\tau}_0+\hat{\tau}_z)/2\nonumber \\
&&+M_x\mathbf{(r)}(\cos\phi\hat{\sigma}_{x}-\sin\phi\hat{\sigma}_{y})(\hat{\tau}_0-
\hat{\tau}_z)/2\nonumber \\
&&+M_y\mathbf{(r)}(\cos\phi\hat{\sigma}_{y}+\sin\phi\hat{\sigma}_{x})(\hat{\tau}_0-\hat{\tau}_z)/2\nonumber \\
&&+\frac{\hat{\sigma}_{z}}{2}(\delta\hat{\tau}_{z}+\delta_2\hat{\tau}_{0})+\Omega_{\bot}\hat{\sigma}_{x}\hat{\tau}_x+\mathcal{U}(\mathbf{r}), \end{eqnarray}
where $M$ is the atomic mass, $\phi$ is the tunable magnetic flux by changing the distance $d$ of bilayer lattice, $\delta_2$ is the effective two-photon detuning, and ${M_x{\mathbf{(r)}}}=\Omega\sin(k_Lx)\cos(k_Ly)$ and $M_y\mathbf{(r)}=\Omega\cos(k_Lx)\sin(k_Ly)$ are nontrivial staggered
spin flips on the $xy$ plane with $\Omega=-\Omega_{\pi}\Omega_{\sigma}/\Delta$.  Here, $\hat{\sigma}_{x,y,z}$ ($\hat{\tau}_{x,y,z}$) are Pauli matrices and $\hat{\sigma}_{0}$ ($\hat{\tau}_{0}$) is identity matrix  acting on the spin (sublattice) space. After applying the gauge transformation ${\cal R}=(\hat{\sigma}_0\hat{\tau}_0+\hat{\sigma}_0\hat{\tau}_z)/2+(\cos\frac{\phi}{2}\hat{\sigma}_x-\sin\frac{\phi}{2}\hat{\sigma}_y)(\hat{\sigma}_0\hat{\tau}_0-\hat{\sigma}_0\hat{\tau}_z)/2
$, the Hamiltonian (\ref{Eq.realspace}) is reduced to
\begin{eqnarray}\label{single}
{\boldsymbol{h}}=&&\frac{{\mathbf{p}}^{2}}{2M}+M_x\mathbf{(r)}\hat{\sigma}_{x}\hat{\tau}_0+
M_y\mathbf{(r)}\hat{\sigma}_{y}\hat{\tau}_z+\frac{\hat{\sigma}_{z}}{2}(\delta\hat{\tau}_{0}+\delta_2\hat{\tau}_{z})\nonumber \\
&&+\Omega_{\bot}(\cos\frac{\phi}{2}\hat{\sigma}_{0}\hat{\tau}_x+\sin\frac{\phi}{2}\hat{\sigma}_{z}\hat{\tau}_y) +\mathcal{U}(\mathbf{r}).
\end{eqnarray}

For applying sufficiently strong lattice potential, the tight-binding Hamiltonian considering only the nearest-neighbor hopping for the lowest $s$ orbit is given by (see Appendix~\ref{appA} for more details)
\begin{eqnarray} \label{realspacelat}
{\cal{H}}={\cal{H}}_+ + {\cal{H}}_- + {\cal{H}}_{\bot},
\end{eqnarray}
where ${\cal{H}}_{\pm}$ represents the lattice Hamiltonian for layer A (+) and B (-) and ${\cal{H}}_{\bot}$ is the Raman-assisted spin-flip hopping between layer A and B. After some mathematical derivations, the lattice Hamiltonian reads
\begin{eqnarray}\label{latticeHa}
{\cal{H}}_{\pm}=-&&it_0\sum_{\mathbf{j}}(\hat{\psi}^{\dag}_{\mathbf{j},\pm}\hat{\sigma}_{x}\hat{\psi}_{\mathbf{j}+\mathbf{1}_x,\pm} \pm\hat{\psi}^{\dag}_{\mathbf{j},\pm}\hat{\sigma}_{y}\hat{\psi}_{\mathbf{j} +\mathbf{1}_y,\pm})+ {\rm H.c.}
\nonumber \\
-&&t\sum_{\mathbf{j}}(\hat{\psi}^{\dag}_{\mathbf{j},\pm}\hat{\sigma}_{z}\hat{\psi}_{\mathbf{j}+\mathbf{1}_x,\pm}+ \hat{\psi}^{\dag}_{\mathbf{j},\pm}\hat{\sigma}_{z}\hat{\psi}_{\mathbf{j}+\mathbf{1}_y,\pm})+{\rm H.c.}
\nonumber \\
+&&\frac{\delta\pm\delta_2}{2}\sum_{\mathbf{j}}\hat{\psi}^{\dag}_{\mathbf{j},\pm}\hat{\sigma}_{z}\hat{\psi}_{\mathbf{j},\pm},
\nonumber \\
{\cal{H}}_{\bot} =&&t_{\bot}\sum_{\mathbf{j}}(e^{-i\phi/2}\hat{a}^{\dag}_{\mathbf{j},\uparrow}\hat{b}_{\mathbf{j},\uparrow}+e^{i\phi/2}\hat{a}^{\dag}_{\mathbf{j},\downarrow}\hat{b}_{\mathbf{j},\downarrow})+{\rm H.c.}, \nonumber
\end{eqnarray}
whose corresponding schematic is depicted in Figs. \ref{model}(b) and \ref{model}(c). Here $\hat{a}^{\dag}_{\mathbf{j},\sigma}$ ($\hat{b}^{\dag}_{\mathbf{j},\sigma}$) is the annihilate operator for spin-$\sigma$ atoms of layer $A$ ($B$) in the $\mathbf{j}=(m,n)$ site with $\hat{\psi}_{\mathbf{j},+}=[\hat{a}_{\mathbf{j},\uparrow},\hat{a}_{\mathbf{j},\downarrow}]^T$ and $\hat{\psi}_{\mathbf{j},-}=[\hat{b}_{\mathbf{j},\uparrow},\hat{b}_{\mathbf{j},\downarrow}]^T$, $t$ ($t_{\bot}$) is the intralayer (interlayer) spin-independent hopping matrix element, and $t_{0}$ is the matrix element for Raman-assisted spin-flip hopping. For convenience, we introduce two lattice unit vectors $\mathbf{1}_x=(1,0)$ and $\mathbf{1}_y=(0,1)$. To obtain Hamiltonian~(\ref{realspacelat}), a gauge transformations $\hat{a}_{\mathbf{j},\downarrow}\rightarrow i(-1)^{(m+n)}\hat{a}_{\mathbf{j},\downarrow}$ and $\hat{b}_{\mathbf{j},\downarrow}\rightarrow i(-1)^{(m+n)}\hat{b}_{\mathbf{j},\downarrow}$ are employed to eliminate the staggered facto for the Raman-induced spin-flip hopping process~\cite{PhysRevA.95.023611}.

In momentum space under periodic boundary condition, the Hamiltonian (\ref{realspacelat}) after performing Fourier transformation is given by
\begin{eqnarray}\label{kspace}
\mathcal{H}(\mathbf{k})=&&d_{x0}(\mathbf{k})\hat{\sigma}_{x}\hat{\tau}_0+d_{yz}(\mathbf{k})\hat{\sigma}_{y}\hat{\tau}_z+d_{z0}(\mathbf{k})\hat{\sigma}_{z}\hat{\tau}_0\nonumber \\
&&+\frac{\delta_2}{2}\hat{\sigma}_{z}\hat{\tau}_z+t_{\bot}(\cos\frac{\phi}{2}\hat{\sigma}_{0}\hat{\tau}_x+\sin\frac{\phi}{2}\hat{\sigma}_{z}\hat{\tau}_y),
\end{eqnarray}
where $\mathbf{k}=(k_x, k_y)$ is in the first Brillouin zone (FBZ), $d_{x0}(\mathbf{k})=2t_0\sin(k_xa)$, $d_{yz}(\mathbf{k})=2t_0\sin(k_ya)$, and $d_{z0}(\mathbf{k})=\frac{\delta}{2}-2t\cos(k_xa)-2t\cos(k_ya)$. Such a system of Eq.~(\ref{kspace}) preserves the particle-hole symmetry ${\cal C}\mathcal{H}(\mathbf{k}){\cal C}^{-1}=-\mathcal{H}(-\mathbf{k})$, where ${\cal C}=\hat{\sigma}_x\hat{\tau}_z\mathcal{K}$ with $\mathcal{K}$ being the complex conjugate operator. Then the system belongs to the symmetry class \rm{D}, where its boundary states could host some properties analogous to the Majorana modes in topological superconductors~\cite{RevModPhys.87.137}.

Furthermore, we find that the system also preserves the inversion symmetry $\mathcal{P}\mathcal{H}(\mathbf{k})\mathcal{P}^{-1}=\mathcal{H}(-\mathbf{k})$ with $\mathcal{P}=\hat{\sigma}_z\hat{\tau}_0$. It will support a topologically protected SOTPs with existing corner states. Noteworthily, the Hamiltonian (\ref{kspace}) possesses ${\cal P}$-symmetry could provide a simple detection method to readily distinguished the different topological phases as discussed in Sec~\ref{Sec3}. In the absence of two-photon detuning $\delta_2=0$, we further note that a time-reversal symmetry $\mathcal{T}\mathcal{H}(\mathbf{k})\mathcal{T}^{-1}=\mathcal{H}(-\mathbf{k})$ appears with $\mathcal{T}=\hat{\sigma}_z\hat{\tau}_x\mathcal{K}$. Therefore, the emerged topological phases are associated with $\mathcal{P}\mathcal{T}$ symmetry with $(\mathcal{P}\mathcal{T})^2=1$, which have been attracted much attention in recent years corresponding to many exotic quantum phenomena in the twisted bilayer graphene~\cite{PhysRevLett.121.126402,PhysRevLett.123.036401,PhysRevB.100.195135,PhysRevX.9.021013,PhysRevB.99.045140,Ahn_2019,PhysRevLett.124.167002,PhysRevLett.125.053601,PhysRevLett.125.126403,PhysRevLett.126.027002}.

\begin{table}[!htbp]
\caption{The expressions of chiral-mirror symmetry operator $\mathcal{\widetilde{M}}_{\alpha}$ for four different magnetic flux configurations. The subscript $\alpha=[x,y,+.-]$ denotes the direction of chiral-mirror line as described in the text.}
\label{symme}
\begin{tabular}{|l|l|l|}
\hline
Flux  &Chiral-mirror symmetry & Chiral-mirror symmetry \\
\hline
$\phi=\frac{\pi}{2}$   & $\mathcal{\widetilde{M}}_-=(\hat{\sigma}_x\hat{\tau}_z+\hat{\sigma}_y\hat{\tau}_0)/\sqrt{2}$ & $\mathcal{\widetilde{M}}_+=(\hat{\sigma}_x\hat{\tau}_0-\hat{\sigma}_y\hat{\tau}_z)/\sqrt{2}$
\\
\hline
$\phi=\pi$   &  $\mathcal{\widetilde{M}}_x=\hat{\sigma}_y\hat{\tau}_0$ &$\mathcal{\widetilde{M}}_y=\hat{\sigma}_x\hat{\tau}_0$
\\
\hline
$\phi=\frac{3\pi}{2}$  & $\mathcal{\widetilde{M}}_+=(\hat{\sigma}_x\hat{\tau}_z-\hat{\sigma}_y\hat{\tau}_0)/\sqrt{2}$ & $\mathcal{\widetilde{M}}_-=(\hat{\sigma}_x\hat{\tau}_0+\hat{\sigma}_y\hat{\tau}_z)/\sqrt{2}$
\\
\hline
$\phi=2\pi$  & $\mathcal{\widetilde{M}}_y=\hat{\sigma}_x\hat{\tau}_z$  & $\mathcal{\widetilde{M}}_x=\hat{\sigma}_y\hat{\tau}_z$
\\
\hline
\end{tabular}
\end{table}

For magnetic flux $\phi=n\pi/2$ with $n$ being the integer, the system also possesses two different chiral-mirror symmetries satisfying the anti-commutations relation with ${\bf k}$-space Hamiltonian, $[\mathcal{\widetilde{M}}_{\alpha}, \mathcal{H}(\mathbf{k})]_+=0$. The subscript of $\mathcal{\widetilde{M}}_{\alpha}$ represents the direction of chiral-mirror symmetry line, e.g. the subscript $\alpha=+$ ($-$) denotes the symmetry line direction along $\vec{e}_x + \vec{e}_y$ ($\vec{e}_x-\vec{e}_y$). Explicitly, the operators of chiral-mirror symmetries are given in Table~\ref{symme}. For an odd number of $n$ (one-quarter and three-quarter integer magnetic fluxes), the chiral-mirror symmetry operators $\mathcal{\widetilde{M}}_+$ and $\mathcal{\widetilde{M}}_-$ satisfy $\mathcal{\widetilde{M}}_+\mathcal{H}(k_x,k_y)\mathcal{\widetilde{M}}_+^{-1}=-\mathcal{H}(k_y,k_x)$ and $\mathcal{\widetilde{M}}_-\mathcal{H}(k_x,k_y)\mathcal{\widetilde{M}}_-^{-1}=-\mathcal{H}(-k_y,-k_x)$. As for an even number of $n$ (half-integer and integer magnetic fluxes), the chiral-mirror symmetry operators $\mathcal{\widetilde{M}}_x$ and $\mathcal{\widetilde{M}}_y$ satisfy $\mathcal{\widetilde{M}}_x\mathcal{H}(k_x,k_y)\mathcal{\widetilde{M}}_x^{-1}=-\mathcal{H}(k_x,-k_y)$ and $\mathcal{\widetilde{M}}_y\mathcal{H}(k_x,k_y)\mathcal{\widetilde{M}}_y^{-1}=-\mathcal{H}(-k_x,k_y)$. For random integer $n$ of magnetic flux, the product of two chiral-mirror symmetry operators is emerged to the inversion symmetry operator $\mathcal{P}$, e.g, $i\mathcal{\widetilde{M}}_x\mathcal{\widetilde{M}}_y={\cal P}$ and $i\mathcal{\widetilde{M}}_-\mathcal{\widetilde{M}}_+={\cal P}$. In addition, we should note that the chiral-mirror symmetry of system is broken for the non-quantized magnetic flux ($\phi\neq n\pi/2$). Without loss of generality, we will focus on investigation of the quantized magnetic flux ($\phi=n\pi/2$) in the following.

\begin{figure}[!htp]
\includegraphics[width=1\columnwidth]{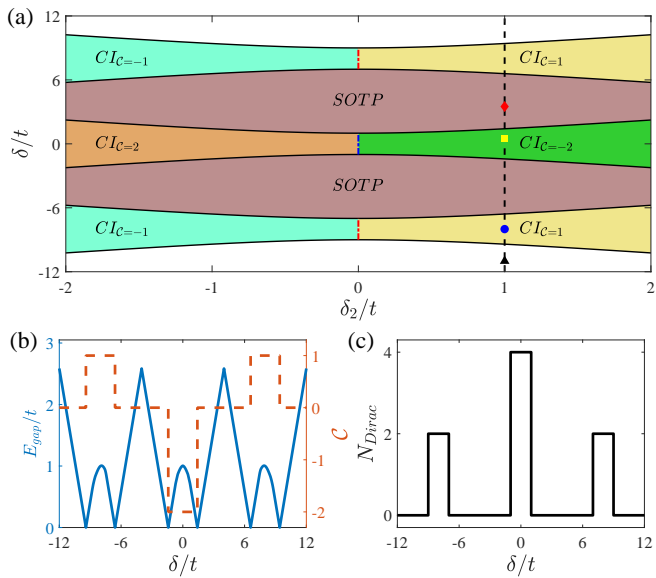}
\caption{\label{diagram}(color online). (a) The magnetic flux-independent phase diagram in the $\delta_2$-$\delta$ parameter plane. The black solid lines denote the phase boundaries of topological phase transitions. And the red (blue) dotted line corresponds to the gapless Dirac semimetals with hosting two (four) Dirac points. The dashed line with $\delta_2/t=1$ is guides to the eye. (b) $\delta$ dependence of energy gap $E_{\rm gap}$ (solid line) and Chern number (dashed line) with $\delta_2/t=1$. (c) $\delta$ dependence of the number of zero-energy Dirac points with $\delta_2=0$. In (a)-(c), the other parameters are $t_{\perp}/t=0.5$ and $t_0/t=1$.}
\end{figure}

\section{Topological Phases Diagram\label{Sec3}}
To characterize the topology of system, we calculate the
energy spectrum by diagonalizing Hamiltonian (\ref{kspace}), $\mathcal{H}(\mathbf{k}) |\psi_{n}({\bf k})\rangle = E_{n}({\bf k})|\psi_{n}({\bf k})\rangle$,
where $ E_{n}({\bf k})$ ($|\psi_{n}({\bf k})\rangle$) denotes the eigenenergies (eigenstates) with $n=\{1, 2, 3, 4\}$ indexing the helicity branches. Thus the direct bulk gap between the two middle branches, $E_{\rm gap}=\min[E_{3}({\bf k})-E_{2}({\bf k})]$, can be obtained. For the conventional topological states, the topological properties of the system are characterized by the first Chern number computed using the two lower occupied branches with $n=1$ and $2$~\cite{RevModPhys.88.035005}.

Figure~\ref{diagram} (a) shows the phase diagram in the $\delta_2$-$\delta$ parameter plane. Here, a gapped Chern insulator with a nonzero Chern number (${\cal C}\neq 0$) is denoted by ``CI$_{\mathcal{C}}$''. A gapped phase with a zero Chern number (${\cal C}=0$) but hosting 0D topologically protected boundary states is denoted by ``SOTP''. Remarkably, the Chern insulators of  ``CI$_{\mathcal{C}=\pm1}$'' and ``CI$_{\mathcal{C}=\pm2}$'' with possessing singly ($|{\mathcal{C}}|=1$) and doubly ($|{\mathcal{C}}|=2$) topological charges are realized. Analytically, the topological phase transitions corresponding to the vanishing bulk gap ($E_{\rm gap}=0$) satisfy the conditions
\begin{align}
\delta=\pm2\sqrt{\delta_2^2/4+t_{\perp}^2}~ {\rm and} ~\delta=\pm8 t \pm2\sqrt{\delta_2^2/4+t_{\perp}^2},
 \end{align}
which divide the $\delta_2$-$\delta$ parameter plane into six regions associated with the different value of ${\cal C}$, as shown in  Fig.~\ref{diagram} (a). Although these curves corresponding to the topological phase transitions is independent of $\phi$, the magnetic flux will dominate the position of topological boundary states for SOTP, as we shall see below.

To understand these phases, we plot the energy gap $E_{\rm gap}$ (solid line) and Chern number (dashed line) as functions of the Zeeman
shift $\delta$ with fixing two-photon detuning $\delta_2/t=1$ in Fig.~\ref{diagram}(b). As can be seen, the sudden changes of the Chern number is associated with the bulk gap closed and reopened at the phase boundaries of topological phase transitions. We note that CI phases only exist for nonzero two-photon detuning ($\delta_2\neq0$) with breaking $\mathcal{P}\mathcal{T}$ symmetry. In fact, the Chern number for gapped band structures must be zero (${\cal C}=0$) under $\mathcal{P}\mathcal{T}$ symmetry~\cite{Ahn_2019}. Furthermore, the phase boundaries between CI phases with opposite Chern number indeed define the gapless Dirac semimetals at $\delta_2=0$, which exhibits $1$D boundary states protected by the $\mathcal{P}\mathcal{T}$ symmetry. Clearly, the Dirac semimetals with two (four) band degenerated points (Dirac points) is observed in Fig.~\ref{diagram}(c), corresponding to the topological phase transition between CI phases with $\mathcal{C}=\pm1$ ($\mathcal{C}=\pm2$). In addition, we should note that the system is not necessarily a topological trivial phase due to existing higher-order topological states even in a region with ${\cal C}=0$. Interestingly, a gapped SOTP ($\mathcal{C}=0$) possessing topologically protected $0$D boundary states can be reduced in contrast to the gapped CI phase (${\cal C}\neq0$) and gapless Dirac semimetals both of which host $1$D boundary states.

To gain more insight into the topological properties for SOTP, we calculate the Wilson loop spectra of system~\cite{PhysRevLett.52.2111,PhysRevLett.48.359,PhysRevLett.62.2747}, which can be used to readily distinguish SOTP and topological trivial phase (${\cal C}=0$). More specifically, we introduce the Wilson loop operator along $y$-direction in the $2$D FBZ
\begin{eqnarray}\label{Eq.loop}
W_{(k_x,k_{0y}\rightarrow k_{0y}+2\pi)}=\lim_{N\rightarrow\infty}F_0F_1...F_{N-2}F_{N-1},
\end{eqnarray}
where $k_{0y}$ is the arbitrary initial point and $F_j$ is the overlap matrix with the element $[F_j]_{mn}=\langle \psi_{m}(k_x,k_{0y}+2\pi j/N)|\psi_{n}(k_x,k_{0y}+2\pi(j+1)/N)\rangle$ and $m,n$ being the indexes of two lower occupied bands. By diagonalizing the Wilson loop operator, the $k_{0y}$-independent eigenvalue takes the form as
\begin{eqnarray}\label{Eq.loopeigen}
W_{(k_x,k_{0y}\rightarrow k_{0y}+2\pi)}|\psi_n\rangle=e^{i\theta_n(k_x)}|\psi_n\rangle.
\end{eqnarray}
The phase factor $\theta_n(k_x)$ identifies the center of the Wannier states, which can completely characterize the different topological nontrivial phases~\cite{Book_short}.

\begin{figure}[!htp]
\includegraphics[width=1\columnwidth]{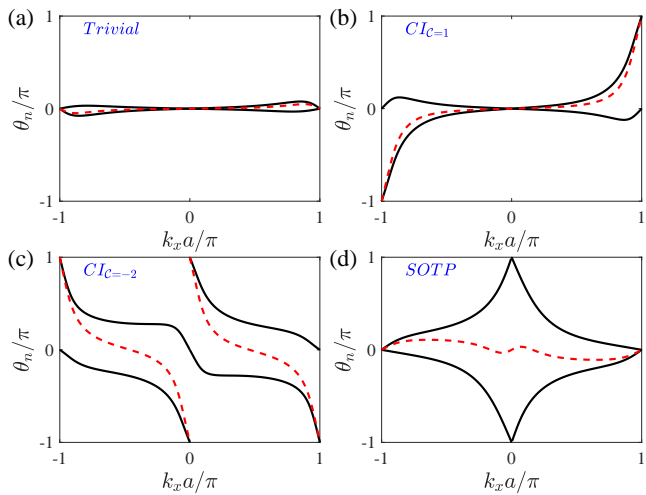}
\caption{\label{wilson}(color online). (a)-(d) The Wilson loop spectra (black solid lines) in occupied subspace and their sum (red dashed line) for different value of $\delta/t=-11$, $-8$, $0.5$, and $3.5$, respectively. The corresponding phase is denoted, which is indexed by black triangle, blue circle, yellow square and red diamond in Fig.~\ref{diagram}(a), respectively. In (a)-(d), the other parameters are $t_{\perp}/t=0.5$, $t_0/t=1$, $\delta_2/t=1$, and $\phi=\pi/2$.}
\end{figure}

Figures~\ref{wilson} shows the typical Wilson loop spectra for different gapped phases. The sum of Wilson spectra $\theta\equiv\theta_1+\theta_2$, shows a flow structure in proportion to the Chern number ${\cal C}$ for the gapped CI phase~\cite{Book_short}. We can visualize Wilson spectra $\theta$ for a topological nontrivial phase characterized by the spin texture with $2\pi$ and $-4\pi$ phase winding, corresponding to the CI phase with $\mathcal{C}=1$ and $\mathcal{C}=-2$, as illustrated in Figs.~\ref{wilson}(b) and~\ref{wilson}(c), respectively. Clearly, Figs.~\ref{wilson}(a) and~\ref{wilson}(d) both display the vanishing flow of $\theta$ associating with zero Chern number (${\cal C}=0$) for topological trivial phase and SOTP. Interestingly, SOTP exhibits a nontrivial winding Wilson loop spectrum with crossing points at $\theta_1=\theta_2=0$ and $\theta_1=\theta_2=\pi$ [Fig.~\ref{wilson}(d)], which is in contrast to the topological trivial phase with crossing points only appearing at $\theta_1=\theta_2=0$ [Fig.~\ref{wilson}(a)].

\begin{table}[!htbp]
\caption{The eigenvalues of $\mathcal{P}$ operator at high-symmetry points and corresponding to the Wilson loop spectra for the same parameters as in Figs.~\ref{wilson}(a)-\ref{wilson}(d).}
\label{Ieigen}
\begin{tabular}{|l|l|l|l|l|}
\hline
phase                  &trivial &$CI_{\mathcal{C}=1}$  &$CI_{\mathcal{C}=-2}$  &SOTP   \\
\hline
$p_n(0,0)$             &$+1,+1$   &$+1,+1$            &$+1,+1$             &$+1,+1$  \\
\hline
$p_n(0,\pi/a)$         &$+1,+1$   &$+1,+1$            &$-1,+1$             &$-1,-1$   \\
\hline
$\theta_n(k_x=0)$      &$0,0$   &$0,0$            &$0,\pi$           &$\pi,\pi$  \\
\hline
$p_n(\pi/a,0)$         &$+1,+1$   &$+1,+1$            &$-1,+1$             &$-1,-1$  \\
\hline
$p_n(\pi/a,\pi/a)$     &$+1,+1$   &$-1,+1$            &$-1,-1$             &$-1,-1$  \\
\hline
$\theta_n(k_x=\pi/a)$  &$0,0$   &$0,\pi$          &$0,\pi$           &$0,0$   \\
\hline
\end{tabular}
\end{table}

\begin{figure}[!htp]
\includegraphics[width=1\columnwidth]{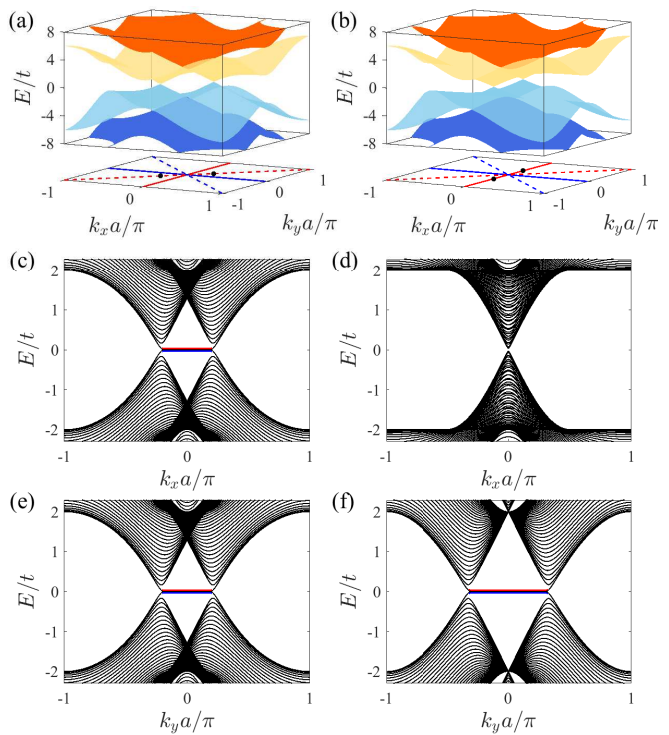}
\caption{\label{Dirac}(color online). (a)-(b) The energy band of Dirac semimetals as a function of $k_x$ and $k_y$ for different magnetic flux $\phi$. Here the black dots indicate the location of Dirac points. The red (blue) dashed line denotes the diagonal (off-diagonal) line with $k_x=k_y$ ($k_x=-k_y$). (c)-(d) [(e)-(f)] The corresponding band spectra of Dirac semimetals under the hard-wall confinement with OBC along $y$- ($x$-) direction for different $\phi$. Blue and red solid lines denote the zero-energy edge states locating at the opposite end. In (a)-(f), the other parameters are $t_{\perp}/t=2$, $t_0/t=1$, $\delta/t=8$, and $\delta_2=0$. From the left to the right columns, the magnetic flux is $\phi={\pi}/{2}$ and $\pi$, respectively.}
\end{figure}

Remarkably, we find that the pattern of Wilson loop spectra are protected by the inversion symmetry $\mathcal{P}$. Under $\mathcal{P}$-symmetry, the eigenstates at high-symmetry points must be the eigenstates of inversion symmetry operator, $\mathcal{P}|\psi_n\rangle=p_n|\psi_n\rangle$ with $p_n=\pm1$. These eigenvalues determine the Wilson loop spectra at the symmetric momentum $\theta_n({k_x\in\{0,\pi/a\}})$~\cite{PhysRevB.96.245115,PhysRevB.89.155114}, as displayed in Table.~\ref{Ieigen}. Specifically, the eigenvalues at all four high-symmetry points satisfy $p_n=+1$, associating with $\theta_n({k_x\in\{0,\pi/a\}})=0$ for the topological trivial phase shown in the second collum of Table.~\ref{Ieigen}. As to the CI phase with $\mathcal{C}=1$, one of the eigenvalues $p_n$ at $(\pi/a,\pi/a)$ will change its sign corresponding to the third collum of Table.~\ref{Ieigen}. Therefore, one of $\theta_n(\pi/a)$ changes its value from $0$ to $\pi$, corresponding to one of the occupied bands acquiring $\pi$ Berry phase along the line $k_x=\pi/a$~\cite{Songeaao4748,PhysRevA.102.013301}. Moreover, the similar results are observed for CI phase with $\mathcal{C}=2$ along the line $k_x=0$, where both $\theta_n(\pi/a)$ and $\theta_n(0)$ will change their values from $0$ to $\pi$ as listed in the fourth collum of Table.~\ref{Ieigen}.

For SOTP, two eigenvalues of $p_{n}$ exhibit the opposite sign at high-symmetric points along $k_x=0$, despite its values own the same sign at high-symmetry points along $k_x=\pi/a$. As a result, the Wilson loop spectra satisfy $\theta_n(0)=\pi$ and $\theta_n(\pi/a)=0$, as displayed in Fig.~\ref{wilson}(d). We should note that the Wilson loop spectra for SOTP are closely related  to the emergence of $0$D corner states. As to the experimental feasibility, the different types of gapped topological phases including CI phases and SOTP with unique structure of Wilson loop spectra can be readily distinguished by measuring spin texture at high-symmetry points of the system via the spin-resolved time-of-flight imaging~\cite{SCPJW2016SOC,Songeaao4748}.

\section{First-order topological phase\label{Sec4}}

\begin{figure}[!htp]
\includegraphics[width=1\columnwidth]{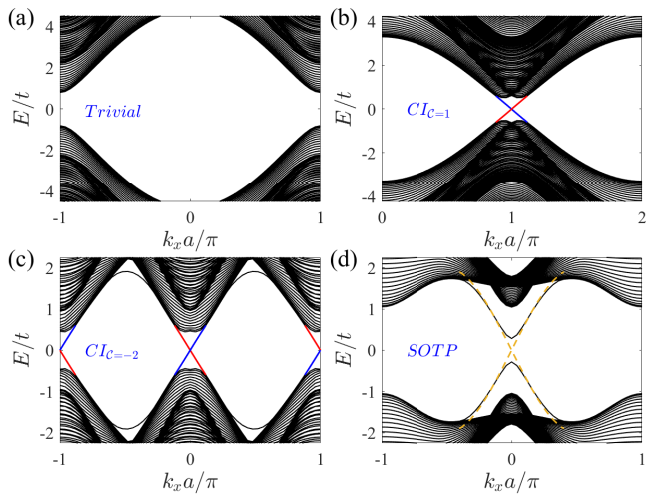}
\caption{\label{diagram_energy_states}(color online). (a)-(d) The energy spectra under the hard-wall confinement with OBC along $y$-direction for $\delta/t=-11, -8, -0.5$, and $3.5$, respectively. The corresponding phase is denoted by black triangle, blue circle, yellow square, and red diamond in Fig.~\ref{diagram}(a). In (b)-(c), the $1$D boundary states with possessing the opposite chirality locating at the bulk gap are characterized by red and blue lines. In (d), the yellow dashed line denotes two degenerated counter-propagating chiral edge states for the uncoupled bilayer lattice with $t_{\perp}/t=0$. In (a)-(d), the other parameters are $t_0/t=1$, $\delta_2/t=1$, $t_{\perp}/t=0.5$, and $\phi={\pi}/{2}$.}
\end{figure}

In order to further reveal the topologies of Dirac semimetals and CI phases, we calculate the topological protected $1$D boundary states according to the conventional bulk-boundary correspondence. In particular, we find that the boundary states for Dirac semimetals can be directly manipulated by the magnetic flux $\phi$. Figures.~\ref{Dirac}(a) and \ref{Dirac}(b) displays the typical band structure of Dirac semimetals for the magnetic flux $\phi={\pi}/{2}$ and $\pi$, respectively. As can be seem, two Dirac points with doubly degeneracies are observed when $\delta_2=0$. We should emphasize that these Dirac points are topological nontrivial, i.e. a closed path around one Dirac point in FBZ can give rise to a nontrivial Berry phase $\pi$. Although the boundaries of topological phase transitions [Fig.~\ref{diagram} (a)] and the number of Dirac points [Fig.~\ref{diagram} (c)] are independent of $\phi$, the position of two Dirac points can be highly controlled by the magnetic flux $\phi$. For the quantized magnetic flux ($\phi= n\pi/2$), the Dirac points will locate at the symmetric line of the chiral-mirror symmetry in Table~\ref{symme}. Moreover, we also check that the ${\bf k}$-space distance between the two Dirac points is also dependent on the value of $\phi$.

To process further, an important property of Dirac semimetals is the existence of exhibiting {$\cal P T$-symmetry} protected boundary states, which link a pair of Dirac points in the surface Brillouin zone. For $\phi={\pi}/{2}$, the $1$D boundary states with lining two Dirac points are both appeared by imposing a hard-wall
confinement along the $y$- [Fig.~\ref{Dirac}(c)] or $x$- [Fig.~\ref{Dirac}(e)] direction under the open boundary condition (OBC). In contrast, two Dirac points only project into a single point at $k_x=0$ in the surface Brillouin zone expanded by $k_x$ for $\phi=\pi$. Therefore, the $1$D boundary states is vanished under the OBC along $y$-direction, as shown in Fig.~\ref{Dirac}(d). Interestingly, the $1$D boundary states will reappear when the hard-wall confinement is imposed along the $x$-direction [Fig.~\ref{Dirac}(f)]. Similar results can also be observed for the other quantized magnetic flux, e.g., $\phi=3\pi/2$ and $\phi=2\pi$. Finally, we should emphasize that the $\mathcal{P}\mathcal{T}$-symmetry protected $1$D boundary states of Dirac semimetals can be tuned by the interplay of magnetic flux and hard-wall confinement direction of OBC.

\begin{figure*}[!htp]
\includegraphics[width=2\columnwidth]{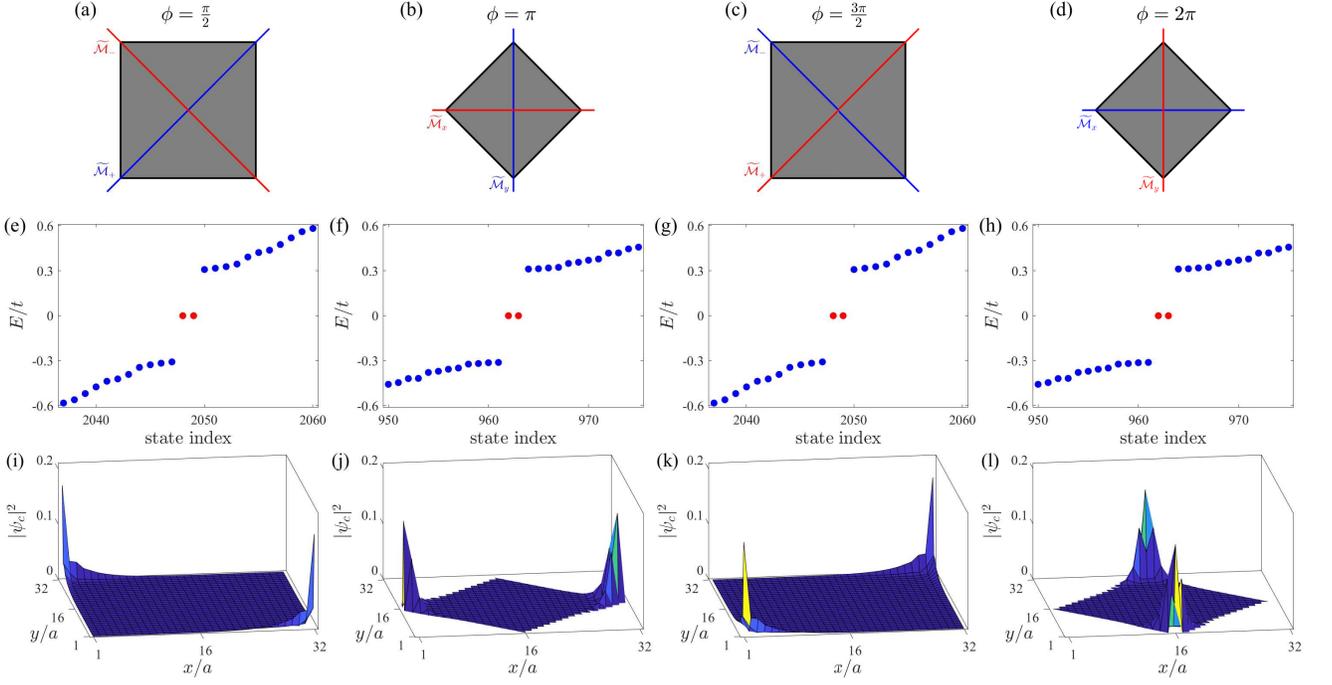}
\caption{\label{all_open}(color online). (a)-(d) The chiral-mirror symmetry lines for different magnetic flux $\phi$, corresponding to the OBCs for generating corner states. Two corner states appear in the boundaries of one certain chiral-mirror symmetry line (red solid line). (e)-(h) The corresponding energy spectra of SOTPs with two zero-energy corner states marked by the red dots. (i)-(l) The typical distributions $|\psi_c(x,y)|^2$ of the two corner states for different magnetic flux $\phi$. In (e)-(h), the other parameters are $t_{\perp}/t=0.5$, $\delta/t=3.5$, and $\delta_2/t=1$.}
\end{figure*}

Compared with the Dirac semimetals, the gapless bands at the Dirac points will be reopened and split with the {${\mathcal{PT}}$}-symmetry breaking ($\delta_2\neq0$), giving rise to a gapped CI phase with a nonzero Chern number. In Figs.~\ref{diagram_energy_states}(a)-\ref{diagram_energy_states}(d), we plot the typical energy spectra for different phases as a function of $k_y$ calculated by the OBC along the $x$-direction. In contrast to the topological trivial phase [Fig.~\ref{diagram_energy_states}(a)], the CI phases with the different Chern numbers support the in-gap $1$D chiral edge states, as shown in Figs.~\ref{diagram_energy_states}(b) and \ref{diagram_energy_states}(c). Clearly, the topological protected single (double) counterpropagating edge modes with the opposite chirality are confirmed for CI phase with ${\cal C}=1$ (${\cal C}=-2$). Remarkably, there is not existing $1$D gapless boundary states in the bulk gap for the SOTP in Fig.~\ref{diagram_energy_states}(d). Unlike the chiral edge states in the CI phases, we note that two degenerated counter-propagating edge states with opposite chirality could appear under the uncoupled bilayer lattice limit, $t_{\perp}/t=0$, which is consistent with the result of Chern number ${\cal C}=0$ for the SOTP.

\section{Second-order topological phase\label{Sec5}}

Now, we turn to study the topological properties of SOTPs, which are essentially different from the conventional first-order topological phases. In addition, due to the chiral-mirror symmetries in Table~\ref{symme}, we shall focus on the quantized magnetic flux ($\phi=n\pi/2$). To investigate the corner states for SOTPs, we calculate the energy spectra under the hard-wall confinement with OBCs along two orthogonal directions. Interestingly, we find that the required OBCs for emerging corner states are associated with the chiral-mirror symmetries, as displayed in Figs.~\ref{all_open}(a)-\ref{all_open}(d). As can be seen, two zero-energy modes are observed for all quantized magnetic fluxes, as shown in Figs.~\ref{all_open}(e)-\ref{all_open}(h). The appearance of two zero-energy edge states are related to the winding Wilson loop spectra shown in Fig.~\ref{wilson}(d). Indeed, the winding Wilson loop spectra indicate that the subsystems for ultracold atoms trapped in bilayer optical lattices hold an opposite topological nontrivial Chern number, $\mathcal{C}_{A}=-\mathcal{C}_{B}=\pm1$. In the uncouple limit, $t_{\perp}/t=0$, two layers of subsystems will support $1$D boundary states with opposite chirality resulting in the conventional bulk-boundary correspondence. In addition, the Raman-assisted spin-flip $t_{\perp}$ between two layers will act as boundary mass terms in system Hamiltonian that describes the dispersion of edge states and gaps out $1$D chiral edge states [Fig.~\ref{diagram_energy_states}(d)]. Interestingly, the effective boundary mass terms will varnish at the one chiral-mirror symmetric lines given in the second column of
Table~\ref{symme}~\cite{PhysRevLett.119.246401,PhysRevLett.119.246402,PhysRevB.97.205135,PhysRevX.9.011012}. Thus the zero-energy corner states, which locate at the junction of  symmetric line $\mathcal{\widetilde{M}}_{\alpha}$ and boundaries of system, can be expected.

Figures~\ref{all_open}(i)-\ref{all_open}(l) display the typical spatial distribution of wave functions for these zero-energy edge modes under the different quantized magnetic flux $\phi=\pi/2, \pi, 3\pi/2$, and $2\pi$, respectively. Obviously, the emergence of topologically protected corner states with respect to zero-energy modes are unambiguously confirmed. Remarkably, the pair of degenerated corner states appear only in one certain chiral-mirror symmetric line [red solid line in Figs.~\ref{all_open}(a)-\ref{all_open}(d)], although the system exists two distinctly chiral-mirror symmetries, simultaneously. To be specific, the corner states emerge at off-diagonal to diagonal corners when $\phi$ changes from $\pi/2$ to $\phi=3\pi/2$, as shown in Figs.~\ref{all_open}(i) and \ref{all_open}(k). As to $\phi=\pi$ ($\phi=2\pi$), the corner states exist at the horizontal (vertical) corners in Fig.~\ref{all_open}(j) [Fig.~\ref{all_open}(l)]. These results demonstrate that the localization of corner states are highly  manipulated in $xy$ plane by tuning the magnetic flux $\phi$. Furthermore, the exotic bulk-boundary correspondence allows us to explore the non-Abelian braiding of Majorana modes~\cite{PhysRevResearch.2.032068,PhysRevResearch.2.043025} with combination of the particle-hole symmetry.

In order to further reveal the property of SOTPs, we analyse the effect of symmetries on the locations of corner states. Obviously, the two chiral-mirror operators commute with the particle-hole symmetry operator, $[\mathcal{\widetilde{M}}_{\alpha}, \mathcal{C}]=0$. In this case, the corner states can appear at both chiral-mirror symmetric lines predicted by Ref.~\cite{PhysRevX.9.011012}. This result can be understood as follows. Without loss of generality, we focus on two-photon detuning $\delta_2=0$. We should emphasize that the bulk gap for SOTPs is always opening when changes $\delta_2=0$ to $\delta_2\neq0$. For ${\delta_2=0}$, our system possesses a chiral symmetry described by the symmetry operator $\mathcal{S}={\cal C}\mathcal{T}=\hat{\sigma}_y\hat{\tau}_y$, and thus falls into symmetry class BDI. Moreover, each chiral-mirror symmetry ($\mathcal{\widetilde{M}}_{\alpha}$) directly corresponds to a mirror symmetry ($\mathcal{{M}}_{\alpha}$) with combination of chiral symmetry ($\mathcal{S}$), e.g., $\mathcal{{M}}_{x}=\mathcal{S}\mathcal{\widetilde{M}}_x=\hat{\sigma}_0\hat{\tau}_y$ and $\mathcal{{M}}_{y}=i\mathcal{S}\mathcal{\widetilde{M}}_y=\hat{\sigma}_z\hat{\tau}_y$ for $\phi=\pi$. We find that $\mathcal{{M}}_{\alpha}$ symmetry associated with $\mathcal{\widetilde{M}}_{\alpha}$ symmetry in the second collum of Table~\ref{symme} commutates with both $\cal C$ and ${\cal T}$ symmetries, $[\mathcal{{M}}_{\alpha}, {\cal C}]=0$ and $[\mathcal{{M}}_{\alpha}, {\cal T}]=0$. In contrast, another one
$\mathcal{{M}}_{\alpha}$ symmetry associated with $\mathcal{\widetilde{M}}_{\alpha}$ symmetry in the third collum of Table~\ref{symme} anticommutates (commutates) with $\cal C$ (${\cal T}$) symmetry, $[\mathcal{{M}}_{\alpha}, {\cal C}]_+=0$ and $[\mathcal{{M}}_{\alpha}, {\cal T}]=0$. According to the study in Ref.~\cite{PhysRevX.9.011012}, the effective boundary mass terms only varnish at the former chiral-mirror symmetric lines given in the second column of
Table~\ref{symme}~\cite{PhysRevLett.119.246401,PhysRevLett.119.246402,PhysRevB.97.205135,PhysRevX.9.011012}, corresponding to the appearance of two $0$D corner states of SOTP. Finally, we note that the emerged SOTP at quantized magnetic flux can be classified into ``intrinsic" HOTPs associated with spatial-symmetry-protected bulk topology. As for $\phi\neq n\pi/2$, the SOTP belongs to ``extrinsic '' SOTPs with hosting the inversion symmetry even the chiral-mirror symmetry is broken ~\cite{PhysRevB.97.205135,PhysRevX.9.011012,ezawa2020edge,asaga2020boundary,wu2020boundary,tiwari2020chiral,khalaf2021boundary}. Furthermore, an ``extrinsic '' SOTP with two zero-energy corner states for the non-quantized magnetic flux are observed, as discussed in Appendix~\ref{appB}.

\section{Conclusion\label{Sec6}}

Based upon the highly controllable techniques of SO coupling and Raman-assisted tunneling, we explore the realization of symmetry-protected higher-order topological states for implementing synthetic magnetic flux of ultracold atoms trapped in bilayer optical lattices. It has been shown that the interplay of two-photon detuning and Zeeman shift gives rise to CI, $\cal P T$-symmetry protected Dirac semimetals, and $\mathcal{\widetilde{M}}_{\alpha}$-symmetry protected SOTP. In contrast to the first-order topological states with bulk-boundary correspondence, the tunable symmetry-protected $0$D corner states for SOTP are observed by tuning the quantized synthetic magnetic flux. Particularly, we show that the  location of 0D corner states is associated with the one certain   chiral-mirror symmetric line of system. Remarkably, these observed topological phases corresponding to the unique structure of Wilson loop spectra can be readily distinguished by measuring spin texture at high-symmetry points via the spin-resolved time-of-flight imaging~\cite{SCPJW2016SOC,Song20191,Wang271}, which will facilitate the experimental explorations of studying novel higher-order topological states. Finally, our scheme can be further extended to study non-Abelian braiding of Majorana corner states in ultracold atomic gases~\cite{PhysRevResearch.2.032068,PhysRevResearch.2.043025}.

\begin{acknowledgments}
{This work is supported by the National Key R$\&$D Program of China (Grant No. 2018YFA0307500), the NSFC (Grants No. 12135018, No. 11874433, No. 12274473, and No. 12104519), and the Guangdong Basic and Applied Basic
Research Foundation (2020A1515110773).}
\end{acknowledgments}

\appendix

\section{Derivation of the Hamiltonian\label{appA}}

In this section, we derive the Hamiltonian in detail with the level diagram and laser configuration given in Fig.~\ref{model}.
The Hamiltonian about the internal states in this bilayer system reads($\hbar=1$)
\begin{eqnarray}\label{Eq.Ham at}
\boldsymbol{h}_+=&&\omega_{AZ}\hat{a}^{\dag}_{\downarrow}\hat{a}_{\downarrow}+\omega_{a}\hat{a}^{\dag}_{e\uparrow}\hat{a}_{e\uparrow}
+(\omega_{a}+\omega'_{AZ})\hat{a}^{\dag}_{e\downarrow}\hat{a}_{e\downarrow}, \nonumber \\
\boldsymbol{h}_-=&&\omega_{BZ}\hat{b}^{\dag}_{\downarrow}\hat{b}_{\downarrow}+\omega_{a}\hat{b}^{\dag}_{e\uparrow}\hat{b}_{e\uparrow}
+(\omega_{a}+\omega'_{BZ})\hat{b}^{\dag}_{e\downarrow}\hat{b}_{e\downarrow}, \nonumber \\
\boldsymbol{h}=&&\boldsymbol{h}_++\boldsymbol{h}_-,
\end{eqnarray}
where $\hat{a}_{\sigma=\uparrow,\downarrow}$ and $\hat{a}_{e,\sigma=\uparrow,\downarrow}$ ($\hat{b}_{\sigma=\uparrow,\downarrow}$ and $\hat{b}_{e,\sigma=\uparrow,\downarrow}$) are, respectively, the annihilation
operators for ground and excited states in layer $A$ ($B$).
$\omega_{a}$ indicates the atomic transition frequency for ground states to excited states, $\omega_{AZ}$ and $\omega_{BZ}$ ($\omega'_{AZ}$ and $\omega'_{BZ}$) are Zeeman shifts of the ground states (excited states) for layer $A$ and $B$, respectively.
Here $\omega_{AZ}\neq\omega_{BZ}$ and $\omega'_{AZ}\neq\omega'_{BZ}$ because of the magnetic field gradient.

Two $\pi$-polarized standing-wave lasers drive the transition $|\sigma\rangle\leftrightarrow |e_{\sigma}\rangle$ with frequency $\omega_L$ and total Rabi frequency $\mathbf{\Omega_{1}}=\Omega_{\pi} [\sin(k_Lx)\cos(k_Ly)+i\cos(k_Lx)\sin(k_Ly)]$, whose detuning is $\Delta=\omega_a-\omega_L$.
We also add a $\sigma$-polarized plane-wave laser along $z$-direction to drive atomic transition $|0\rangle\leftrightarrow |e_{\uparrow}\rangle$ for layer $A$ ($B$) with frequency $\omega_L+\Delta\omega_L$ and Rabi frequency $\Omega_{\sigma}$ ($\Omega_{\sigma}e^{i\phi}$).
Here the phase difference $\phi=\sqrt{2}k_Ld$ is determined by the distance $d$ between two layers.
Then the Hamiltonian for atom-light system reads
\begin{eqnarray}\label{Eq.Hamin at}
\boldsymbol{h}_{+}=&&\omega_{AZ}\hat{a}^{\dag}_{\downarrow}\hat{a}_{\downarrow}+\omega_{a}\hat{a}^{\dag}_{e\uparrow}\hat{a}_{e\uparrow}
+(\omega_{a}+\omega'_{AZ})\hat{a}^{\dag}_{e\downarrow}\hat{a}_{e\downarrow} \nonumber \\
&&+[e^{i\omega_Lt}(\mathbf{\Omega^*_{1}}\hat{a}^{\dag}_{\uparrow}\hat{a}_{e\uparrow}+\mathbf{\Omega^*_{1}}\hat{a}^{\dag}_{\downarrow}\hat{a}_{e\downarrow})\nonumber \\
&&+\Omega^*_{\sigma}e^{i(\omega_L+\Delta\omega_L)t}\hat{a}^{\dag}_{\downarrow}\hat{a}_{e\uparrow}+H.c.], \nonumber \\
\boldsymbol{h}_{-}=&&\omega_{BZ}\hat{b}^{\dag}_{\downarrow}\hat{b}_{\downarrow}+\omega_{a}\hat{b}^{\dag}_{e\uparrow}\hat{b}_{e\uparrow}
+(\omega_{a}+\omega'_{BZ})\hat{b}^{\dag}_{e\downarrow}\hat{b}_{e\downarrow} \nonumber \\
&&+[e^{i\omega_Lt}(\mathbf{\Omega^*_{1}}\hat{b}^{\dag}_{\uparrow}\hat{b}_{e\uparrow}+\mathbf{\Omega^*_{1}}\hat{b}^{\dag}_{\downarrow}\hat{b}_{e\downarrow})
\nonumber \\
&&+\Omega^*_{\sigma}e^{i(\omega_Lt+\Delta\omega_Lt-\phi)}\hat{b}^{\dag}_{\downarrow}\hat{b}_{e\uparrow}+H.c.], \nonumber \\
\boldsymbol{h}=&&\boldsymbol{h}_++\boldsymbol{h}_-.
\end{eqnarray}

To eliminate the time dependent factors, we define the unitary transformation $\widetilde{\mathcal{U}}_A=e^{-iu_At}$ and $\widetilde{\mathcal{U}}_B=e^{-iu_Bt}$ with
\begin{eqnarray}\label{Eq.Unita}
u_+&&=-\Delta\omega_L\hat{a}^{\dag}_{\downarrow}\hat{a}_{\downarrow}
+\omega_L\hat{a}^{\dag}_{e\uparrow}\hat{a}_{e\uparrow}
+(\omega_L-\Delta\omega_L)\hat{a}^{\dag}_{e\downarrow}\hat{a}_{e\downarrow},
\nonumber \\
u_-&&=-\Delta\omega_L\hat{b}^{\dag}_{\downarrow}\hat{b}_{\downarrow}
+\omega_L\hat{b}^{\dag}_{e\uparrow}\hat{b}_{e\uparrow}
+(\omega_L-\Delta\omega_L)\hat{b}^{\dag}_{e\downarrow}\hat{b}_{e\downarrow}.\nonumber \\
\end{eqnarray}
In this rotating frame, the Hamiltonian reads
\begin{eqnarray}\label{Eq.Rotating}
\boldsymbol{h}_{+}=&&\delta_{A}\hat{a}^{\dag}_{\downarrow}\hat{a}_{\downarrow}+\Delta\hat{a}^{\dag}_{e\uparrow}\hat{a}_{e\uparrow}
+(\Delta+\Delta\omega_L+\omega'_{AZ})\hat{a}^{\dag}_{e\downarrow}\hat{a}_{e\downarrow} \nonumber \\
&&+(\mathbf{\Omega^*_{1}}\hat{a}^{\dag}_{\uparrow}\hat{a}_{e\uparrow}+\mathbf{\Omega^*_{1}}\hat{a}^{\dag}_{\downarrow}\hat{a}_{e\downarrow}
+\Omega^*_{\sigma}\hat{a}^{\dag}_{\downarrow}\hat{a}_{e\uparrow}+H.c.), \nonumber \\
\boldsymbol{h}_{-}=&&\delta_{B}\hat{b}^{\dag}_{\downarrow}\hat{b}_{\downarrow}+\Delta\hat{b}^{\dag}_{e\uparrow}\hat{b}_{e\uparrow}
+(\Delta+\Delta\omega_L+\omega'_{AZ})\hat{b}^{\dag}_{e\downarrow}\hat{b}_{e\downarrow} \nonumber \\
&&+(\mathbf{\Omega^*_{1}}\hat{b}^{\dag}_{\uparrow}\hat{b}_{e\uparrow}+\mathbf{\Omega^*_{1}}\hat{b}^{\dag}_{\downarrow}\hat{a}_{e\downarrow}
+\Omega^*_{\sigma}e^{-i\phi}\hat{b}^{\dag}_{\downarrow}\hat{b}_{e\uparrow}+H.c.), \nonumber \\
\boldsymbol{h}=&&\boldsymbol{h}_++\boldsymbol{h}_-.
\end{eqnarray}
Here $\delta_{A}=\omega_{AZ}+\Delta\omega_L$ and $\delta_{B}=\omega_{BZ}+\Delta\omega_L$ known as the two-photon detuning for
layer $A$ and $B$ respectively.
Then we can get the Heisenberg equations about different components
\begin{eqnarray}\label{Eq.Heisenberg}
i\frac{\partial\hat{a}_{\uparrow}}{\partial t}&=&\mathbf{\Omega^*_{1}}\hat{a}_{e\uparrow}, \nonumber \\
i\frac{\partial\hat{a}_{\downarrow}}{\partial t}&=&\delta_{A}\hat{a}_{\downarrow}+\mathbf{\Omega^*_{1}}\hat{a}_{e\downarrow}+{\Omega^*_{\sigma}}\hat{a}_{e\uparrow}, \nonumber \\
i\frac{\partial\hat{a}_{e\uparrow}}{\partial t}&=&(\Delta-i\gamma/2)\hat{a}_{e\uparrow}+\mathbf{\Omega_{1}}\hat{a}_{\uparrow} +{\Omega_{\sigma}}\hat{a}_{\downarrow}, \nonumber \\
i\frac{\partial\hat{a}_{e\downarrow}}{\partial t}&=&(\Delta+\Delta\omega_L+\omega'_{AZ}-i\gamma/2)\hat{a}_{e\downarrow}+\mathbf{\Omega_{1}}\hat{a}_{\downarrow}, \nonumber \\
i\frac{\partial\hat{b}_{\uparrow}}{\partial t}&=&\mathbf{\Omega^*_{1}}\hat{b}_{e\uparrow}, \nonumber \\
i\frac{\partial\hat{b}_{\downarrow}}{\partial t}&=&\delta_{B}\hat{b}_{\downarrow}+\mathbf{\Omega^*_{1}}\hat{b}_{e\downarrow}+{\Omega^*_{\sigma}}e^{-i\phi}\hat{b}_{e\uparrow}, \nonumber \\
i\frac{\partial\hat{b}_{e\uparrow}}{\partial t}&=&(\Delta-i\gamma/2)\hat{b}_{e\uparrow}+\mathbf{\Omega_{1}}e^{i\phi}\hat{b}_{\uparrow} +{\Omega_{\sigma}}\hat{b}_{\downarrow}, \nonumber \\
i\frac{\partial\hat{b}_{e\downarrow}}{\partial t}&=&(\Delta+\Delta\omega_L+\omega'_{BZ}-i\gamma/2)\hat{b}_{e\downarrow}+\mathbf{\Omega_{1}}\hat{b}_{\downarrow},
 \end{eqnarray}
where $\gamma$ is the spontaneous emission rate for excited states. When the atom-light detuning is sufficient large,  $|\Omega_{\sigma,\pi}/\Delta|\ll1$, $|\delta_{\uparrow,\downarrow}/\Delta|\ll1$ and $|\gamma/\Delta|\ll1$, the exited states can be adiabatically eliminated by setting $i\dot{\hat{a}}_{e\uparrow,e\downarrow}=i\dot{\hat{b}}_{e\uparrow,e\downarrow}=0$, which yields
\begin{eqnarray}\label{Eq.exited}
\hat{a}_{e\uparrow}\approx&&-\frac{\mathbf{\Omega_{1}}\hat{a}_{\uparrow}+{\Omega_{\sigma}}\hat{a}_{\downarrow}}{\Delta}, \nonumber \\
\hat{a}_{e\downarrow}\approx&&-\frac{\mathbf{\Omega_{1}}\hat{a}_{\downarrow}}{\Delta}, \nonumber \\
\hat{b}_{e\uparrow}\approx&&-\frac{\mathbf{\Omega_{1}}\hat{b}_{\uparrow}+{\Omega_{\sigma}}e^{i\phi}\hat{b}_{\downarrow}}{\Delta}, \nonumber \\
\hat{a}_{e\downarrow}\approx&&-\frac{\mathbf{\Omega_{1}}\hat{b}_{\downarrow}}{\Delta}.
\end{eqnarray}
Through substitute Eq.~\ref{Eq.exited} to Eq.~\ref{Eq.Heisenberg}, we obtain the effective Heisenberg equation about ground states:
\begin{eqnarray}\label{Eq.ground}
i\frac{\partial\hat{a}_{\uparrow}}{\partial t}=&&-\frac{1}{\Delta}(|\mathbf{\Omega_{1}}|^2\hat{a}_{\uparrow}+\mathbf{\Omega^*_{1}}{\Omega_{\sigma}}\hat{a}_{\downarrow}), \nonumber \\
i\frac{\partial\hat{a}_{\downarrow}}{\partial t}=&&\delta_{A}\hat{a}_{\downarrow}-\frac{1}{\Delta}[(|\mathbf{\Omega_{1}}|^2+|{\Omega_{\sigma}}|^2)\hat{a}_{\downarrow}+
\mathbf{\Omega_{1}}{\Omega^*_{\sigma}}\hat{a}_{\downarrow}], \nonumber \\
i\frac{\partial\hat{b}_{\uparrow}}{\partial t}=&&-\frac{1}{\Delta}(|\mathbf{\Omega_{1}}|^2\hat{b}_{\uparrow}+\mathbf{\Omega^*_{1}}{\Omega_{\sigma}}e^{i\phi}\hat{b}_{\downarrow}), \nonumber \\
i\frac{\partial\hat{b}_{\downarrow}}{\partial t}=&&\delta_{B}\hat{b}_{\downarrow}-\frac{1}{\Delta}[(|\mathbf{\Omega_{1}}|^2+|{\Omega_{\sigma}}|^2)\hat{b}_{\downarrow}+\mathbf{\Omega_{1}}{\Omega^*_{\sigma}}e^{-i\phi}\hat{b}_{\downarrow}].\nonumber \\
\end{eqnarray}
Then the effective Hamiltonian for ground states can be obtained by
\begin{eqnarray}\label{Eq.groundHa}
{\boldsymbol{h}}=&&M_x\mathbf{(r)}\hat{\sigma}_{x}(\hat{\tau}_0+
\hat{\tau}_z)/2+M_y\mathbf{(r)}\hat{\sigma}_{y}(\hat{\tau}_0+\hat{\tau}_z)/2\nonumber \\
&&+M_x\mathbf{(r)}(\cos\phi\hat{\sigma}_{x}-\sin\phi\hat{\sigma}_{y})(\hat{\tau}_0-
\hat{\tau}_z)/2\nonumber \\
&&+M_y\mathbf{(r)}(\cos\phi\hat{\sigma}_{y}+\sin\phi\hat{\sigma}_{x})(\hat{\tau}_0-\hat{\tau}_z)/2\nonumber \\
&&+\frac{\hat{\sigma}_{z}}{2}(\delta\hat{\tau}_{z}+\delta_2\hat{\tau}_{0}),
\end{eqnarray}
where $\delta\equiv\frac{\delta_{B}-\delta_{A}}{2}=g_F \mu_Bb_0 d$ and $\delta_2\equiv\frac{|\Omega_{\sigma}|^2}{2\Delta}-\frac{\delta_{A}+\delta_{B}}{2}$, $\hat{\sigma}_{x,y,z}$ ($\hat{\tau}_{x,y,z}$) are Pauli matrices and $\hat{\sigma}_{0}$ ($\hat{\tau}_{0}$) is identity matrix acting on the spin (sublattice) space.
Further $M_x(r)=\Omega\sin(k_Lx)\cos(k_Ly)$ and $M_y(r)=\Omega\cos(k_Lx)\sin(k_Ly)$ are position-dependent Rabi coupling with $\Omega=-\Omega_{\pi}\Omega_{\sigma}/\Delta$

To couple these two layers, we add another laser to induce the spin-flipping coupling with Rabi frequency $\Omega_{\bot}$ between two layers
 \begin{eqnarray}\label{Eq.coulpe}
{\boldsymbol{h}}_{\bot}=\Omega_{\bot}[\hat{a}^{\dag}_{\uparrow}\mathbf{(r)}\hat{b}_{\downarrow}\mathbf{(r)}+\hat{a}^{\dag}_{\downarrow}\mathbf{(r)}\hat{b}_{\uparrow}\mathbf{(r)}+{\rm H.c.}].
\end{eqnarray}
Further considering the center-of-mass motion, the Hamiltonian of the atom-light system reads
 \begin{eqnarray}\label{Eq.groundHa}
 {\boldsymbol{h}}=&&\frac{{\mathbf{p}}^{2}}{2M}+M_x\mathbf{(r)}\hat{\sigma}_{x}(\hat{\tau}_0+
\hat{\tau}_z)/2+M_y\mathbf{(r)}\hat{\sigma}_{y}(\hat{\tau}_0+\hat{\tau}_z)/2\nonumber \\
&&+M_x\mathbf{(r)}(\cos\phi\hat{\sigma}_{x}-\sin\phi\hat{\sigma}_{y})(\hat{\tau}_0-
\hat{\tau}_z)/2\nonumber \\
&&+M_y\mathbf{(r)}(\cos\phi\hat{\sigma}_{y}+\sin\phi\hat{\sigma}_{x})(\hat{\tau}_0-\hat{\tau}_z)/2\nonumber \\
&&+\frac{\hat{\sigma}_{z}}{2}(\delta\hat{\tau}_{z}+\delta_2\hat{\tau}_{0})+\Omega_{\bot}\hat{\sigma}_{x}\hat{\tau}_x+\mathcal{U}(\mathbf{r})\hat{\sigma}_{0},
\end{eqnarray}
where $M$ is the atomic mass.
After the gauge transformation: ${\cal R}=(\hat{\sigma}_0\hat{\tau}_0+\hat{\sigma}_0\hat{\tau}_z)/2+(\cos\frac{\phi}{2}\hat{\sigma}_x-\sin\frac{\phi}{2}\hat{\sigma}_y)(\hat{\sigma}_0\hat{\tau}_0-\hat{\sigma}_0\hat{\tau}_z)/2
$, we obtain the real space Hamiltonian given in the main text
 \begin{eqnarray}\label{Eq.realspaceapsup}
{\boldsymbol{h}}=&&\frac{{\mathbf{p}}^{2}}{2M}+M_x\mathbf{(r)}\hat{\sigma}_{x}\hat{\tau}_0+
M_y\mathbf{(r)}\hat{\sigma}_{y}\hat{\tau}_z+\frac{\hat{\sigma}_{z}}{2}(\delta\hat{\tau}_{0}+\delta_2\hat{\tau}_{z})\nonumber \\
&&+\Omega_{\bot}(\cos\frac{\phi}{2}\hat{\sigma}_{0}\hat{\tau}_x+\sin\frac{\phi}{2}\hat{\sigma}_{z}\hat{\tau}_y)+\mathcal{U}(\mathbf{r})\hat{\sigma}_{0}.
\end{eqnarray}

\begin{figure}[!htp]
\includegraphics[width=1\columnwidth]{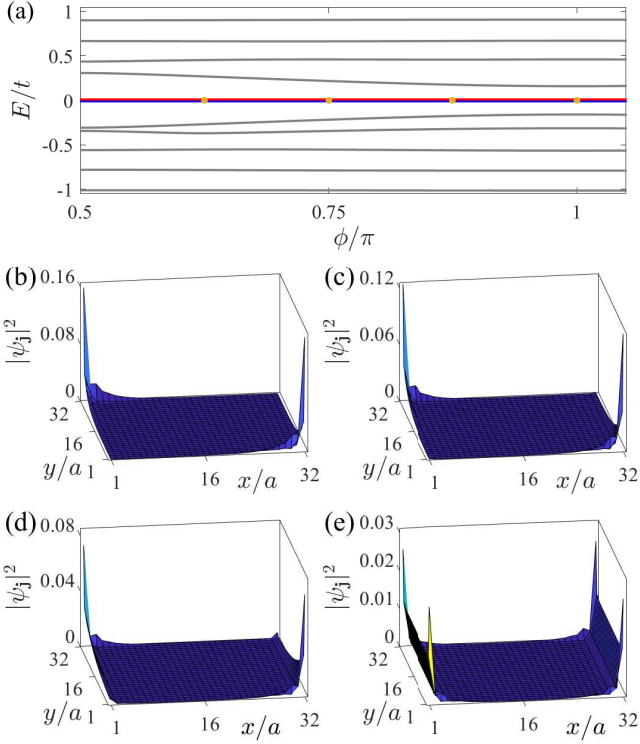}
\caption{\label{other_energy_flux}(color online). (a) The typical spectra with different fluxes for SOTP, where the OBC along $x$- and $y$-directions are set. The red and blue lines index the corner states. (b-e) The distributions of boundary states with $\phi/\pi=0.625$, $0.75$, $0.875$ and $1$, respectively, which are denoted by yellow dots in (a). In all panels, $t_{\perp}/t=0.5$, $t_0/t=1$, $\delta/t=3.5$ and $\delta_2/t=1$ are set.}
\end{figure}

For sufficiently strong lattice potential, the system enters
the tight-binding regime and the corresponding lattice Hamiltonian with the nearest-neighbor hopping between the lowest s orbit is given by:
\begin{eqnarray}\label{Eq.Lattice0}
{\cal{H}_+}=&&~\left[t_0\sum_{\mathbf{j},\zeta=\pm1}(-1)^{(m+n)}\zeta(\hat{a}^{\dag}_{\mathbf{j},\uparrow}\hat{a}_{\mathbf{j}+\zeta\mathbf{1}_x,\downarrow}
-i\hat{a}^{\dag}_{\mathbf{j},\uparrow}\hat{a}_{\mathbf{j}+\zeta\mathbf{1}_y,\downarrow})
\right.\nonumber \\
&&\left. -t\sum_{\mathbf{j},\sigma}\hat{a}^{\dag}_{\mathbf{j},\sigma}\hat{a}_{\mathbf{j}+\mathbf{1}_{x},\sigma}+\hat{a}^{\dag}_{\mathbf{j},\sigma}\hat{a}_{\mathbf{j}+\mathbf{1}_{y},\sigma}+{\rm H.c.}\right]
\nonumber \\
&&+\frac{\delta+\delta_2}{2}\sum_{\mathbf{j}}(\hat{a}^{\dag}_{\mathbf{j},\uparrow}\hat{a}_{\mathbf{j},\uparrow}-\hat{a}^{\dag}_{\mathbf{j},\downarrow}\hat{a}_{\mathbf{j},\downarrow}),
\nonumber \\
{\cal{H}_-}=&&~\left[t_0\sum_{\mathbf{j},\zeta=\pm1}(-1)^{(m+n)}\zeta(\hat{b}^{\dag}_{\mathbf{j},\uparrow}\hat{b}_{\mathbf{j}+\zeta\mathbf{1}_x,\downarrow}
+i\hat{b}^{\dag}_{\mathbf{j},\uparrow}\hat{b}_{\mathbf{j}+\zeta\mathbf{1}_y,\downarrow})
\right.\nonumber \\
&&\left.-t\sum_{\mathbf{j},\sigma}\hat{b}^{\dag}_{\mathbf{j},\sigma}\hat{b}_{\mathbf{j}+\mathbf{1}_{x},\sigma}+\hat{b}^{\dag}_{\mathbf{j},\sigma}\hat{b}_{\mathbf{j}+\mathbf{1}_{y},\sigma}+{\rm H.c.}\right]
\nonumber \\
&&+\frac{\delta-\delta_2}{2}\sum_{\mathbf{j}}(\hat{b}^{\dag}_{\mathbf{j},\uparrow}\hat{b}_{\mathbf{j},\uparrow}-\hat{b}^{\dag}_{\mathbf{j},\downarrow}\hat{b}_{\mathbf{j},\downarrow}),
\nonumber \\
{\cal{H}}=&&{\cal{H}_+}+{\cal{H}_-}\nonumber \\&&+t_{\bot}\sum_{\mathbf{j}}[e^{-i\phi/2}\hat{a}^{\dag}_{\mathbf{j},\uparrow}\hat{b}_{\mathbf{j},\uparrow}+e^{i\phi/2}\hat{a}^{\dag}_{\mathbf{j},\downarrow}\hat{b}_{\mathbf{j},\downarrow}+{\rm H.c.}]
\end{eqnarray}
where $\hat{a}^{\dag}_{\mathbf{j},\sigma}$ ($\hat{b}^{\dag}_{\mathbf{j},\sigma}$) with $\mathbf{j}=(m,n)$ is the annihilate operator in the $m$-($n$-)th site along $x$-($y$-)direction for layer $A$ ($B$), $t$ is the spin-independent hopping matrix element, $t_{\bot}$ is the hopping matrix element between two layers, while $t_{0}$ is the matrix element for Raman-assisted spin-flip hopping, as described in main text.

Obviously, the Raman-assisted nearest-neighbor spin-flip hopping is staggered in Hamiltonian~(\ref{Eq.Lattice0}).
To eliminate the staggered factor, we use the gauge transformation $\hat{a}_{\mathbf{j},\downarrow}\rightarrow (-1)^{(m+n)}i\hat{a}_{\mathbf{j},\downarrow}$ and $\hat{b}_{\mathbf{j},\downarrow}\rightarrow (-1)^{(m+n)}i\hat{b}_{\mathbf{j},\downarrow}$~\cite{PhysRevA.95.023611}, and then the Hamiltonian described in~\eqref{latticeHa} of the main text has been obtained.

\section{The spectra and boundary states with non-quantized fluxes\label{appB}}
To reveal properties of SOTP in our model more comprehensively, we discuss the corner states with general values of magnetic flux in this section.
In this process, the OBC along $x$- and $y$-directions are still retained.
In Figure~\ref{other_energy_flux}(a), we plot the typical energy spectra for SOTP with different fluxes.
As can be seen, the boundary states retain zero-energy, and the energy gap reaches its minimum when $\phi=\pi$.
Actually the energy gap will tend to zero under thermodynamic limit for $\phi=\pi$ with this boundary condition.

\begin{figure}[!htp]
\includegraphics[width=1\columnwidth]{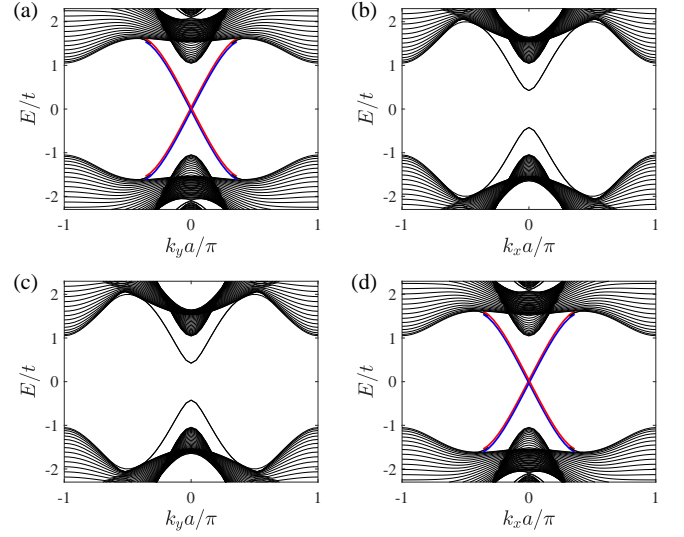}
\caption{\label{sup_energy_oneOBC}(color online). The energy spectrum with flux $\phi=\pi$ (a-b) and $\phi=2\pi$ (c-d). In (a) and (c) [(b) and (d)], the OBC along $x$-direction ($y$-direction) is set. In all panel $t_{\perp}/t=0.5$, $t_0/t=1$, $\delta/t=3.5$ and $\delta_2/t=1$ are set. The gapless $1$D edge states locating at right and left (bottom and up) end are characterized by red and blue lines in (a) [(d)] respectively.}
\end{figure}

Despite boundary states retain zero-energy, their distributions will change along with flux. As discussed in Sec~\ref{Sec5} and shown in Fig.~\ref{all_open}(a) of main text, corner states will locate at off-diagonal corners and are protected by two chiral-mirror symmetries shown in Table~\ref{symme} of main text when $\phi=\pi/2$.
When the magnetic flux deviates from $\pi/2$, these two chiral-mirror symmetries are broken and the corner states will extent along $y$-direction as shown in Figs.~\ref{other_energy_flux}(b-e).
Eventually, these boundary states evolve into $1$D edge states along $(10)$ and $(\overline{1}0)$ ends when $\phi=\pi$ as shown in Fig.~\ref{other_energy_flux}(e).
We emphasis that the edge states along $(01)$ and $(0\overline{1})$ ends are absence under this conditions.
As we increase the magnetic flux continually, the boundary states will shrink around the diagonal corner and are maximally localized when $\phi=3\pi/2$, as displayed in Fig.~\ref{all_open}(c) of main text.
Further the energy gap reaches another minimum when $\phi=2\pi$, where the $1$D edge states appear along $(01)$ and $(0\overline{1})$ ends.

To reveal the properties of $1$D boundary states more deeply, we also plot the energy spectrum with OBC along one direction in Figs.~\ref{sup_energy_oneOBC}.
As demonstrated here, the helical edge states appear (vanish) when $x$- ($y$-) direction takes OBC for $\phi=\pi$, as shown in Figs.~\ref{sup_energy_oneOBC}(a-b).
In contrast, the helical edge states only appear at $(01)$ and $(0\overline{1})$ ends when $\phi=2\pi$, as confirmed by Figs.~\ref{sup_energy_oneOBC}(c-d).

These results confirm the mechanism for inducing corner states.
As discussed in Sec~\ref{Sec5} of the main text, the coupling term between two layers acts as a boundary mass term, which gaps out the edge states and vanishes at certain locations of boundary, giving rise to corner states.
 For $\phi=\pi$ ($\phi=2\pi$), the boundary mass term will vanish along boundaries normal to $x$- ($y$-) direction. Then with the condition shown in Fig.~\ref{all_open}(b) [(d)] of main text, two corner states are generated, while they will become counter-propagating $1$D edge states when $x$- ($y$-) direction takes OBC.
%

%

\end{document}